\documentclass{aastex631}

\begin{document}

\title{{Simulating secondary electron and ion emission from the Cassini spacecraft in Saturn's ionosphere}}

\correspondingauthor{Z. Zhang}
\email{zeqi.zhang17@imperial.ac.uk}

\author[0000-0002-5672-2681]{Z. Zhang}
\affiliation{Blackett Laboratory, Imperial College London, UK}

\author[0000-0002-2015-4053]{R. T. Desai}
\affiliation{Blackett Laboratory, Imperial College London, UK}
\affiliation{Centre for Fusion, Space and Astrophysics, University of Warwick, UK}

\author[0000-0001-9621-211X]{O. Shebanits}
\affiliation{Swedish Institute of Space Physics, Uppsala, Sweden}

\author[0000-0002-5386-8255]{F. L. Johansson}
\affiliation{ESA/ESTEC, Noordwijk, The Netherlands}

\author[0000-0001-6491-1012]{Y. Miyake}
\affiliation{Graduate School of System Informatics, Kobe University, Kobe, Japan}
\author[0000-0001-5846-9109]{H. Usui}
\affiliation{Graduate School of System Informatics, Kobe University, Kobe, Japan}

\begin{abstract}

  The Cassini spacecraft's Grand Finale flybys through Saturn's ionosphere provided unprecedented insight into the composition and dynamics of the gas giant's upper atmosphere and a novel and complex spacecraft-plasma interaction. In this article, we further study Cassini's interaction with Saturn's ionosphere using three dimensional Particle-in-Cell simulations. We focus on understanding how electrons and ions, emitted from spacecraft surfaces due to the high-velocity impact of atmospheric water molecules, could have affected the spacecraft potential and low-energy plasma measurements. The simulations show emitted electrons extend upstream along the magnetic field and, for sufficiently high emission rates, charge the spacecraft to positive potentials. 
  The lack of accurate emission rates and characteristics, however, makes differentiation between the prominence of secondary electron emission and ionospheric charged dust populations, which induce similar charging effects, difficult for Cassini. These results provide further context for Cassini's final measurements and highlight the need for future laboratory studies to support high-velocity flyby missions through planetary and cometary ionospheres. 
\end{abstract}

\section{Introduction}
Cassini's Grand Finale obtained the first ever in-situ measurement of Saturn's ionosphere. Passing through on 22 orbits prior to its final plunge, the spacecraft provided unprecedented observations from inside Saturn's D-ring down to 1,360 km altitude \citep{Ip16,Cravens19,Dougherty18,Hsu18,Lamy18,Mitchell18,Roussos18,Waite18,Wahlund18}. Cassini's Plasma Spectrometer \citep{Young04} was, however, offline post-2012 and significant unknowns remain regarding charged ion and dust populations and their influence on the gas giant's ionosphere. 

Saturn's inner rings are inherently unstable and were identified as raining onto the top of the gas giant's equatorial ionosphere \citep{Connerney84,Northrop82}. This was subsequently observed by Cassini in-situ where the spacecraft's Ion and Neutral Mass Spectrometer (INMS), Cosmic Dust Analyser and Charge Energy Mass Spectrometer 
detected ring fragments consisting of water, silicates and organics in-flowing at estimated fluxes between 4,800 and 45,0000 kg/s \citep{Hsu18,Mitchell18,Waite18}. 
Cassini's high-velocity limited the spectroscoptic plasma measurements to $<$5 u and the composition of Saturn's ionosphere has thus been inferred from the available measurements. For example, Cassini's Radio and Plasma Wave Science antenna observed up to an order of magnitude greater electrons than $1-4$ u positive ions, and  ion populations of $>$ 4 u were therefore inferred to be present \citep{Waite18}. 
Cassini's Langmuir Probe, also, simultaneously measured nearly an order of magnitude greater positive ion currents compared with electron currents \citep{Hadid19,Morooka19,Wahlund18} and 
these observations were thus interpreted as arising from increasingly abundant populations of negatively charged ions and dust with decreasing altitude \citep{Morooka19}, in addition to the larger $>$4 u positive ions. 

Cassini's Langmuir Probe measured the bulk plasma currents and therefore uniquely provides a measure of all ionospheric plasma constituents. As an integral measurement, however, the interpretation of this data-set is non-trivial. Two distinct interpretations of the LP data thus exist within the literature. \citet{Morooka19} first report the apparent current discrepancies as arising from significant populations of charged dust, an interpretation  which has formed the basis for sequential studies of Saturn's ionosphere \citep[e.g.][]{Hadid19,Wahlund18,Shebanits20,Zhang21}. \citet{Johansson22}, however, recently suggested that less dust is present and that the LP current imbalance is caused by secondary electrons and ions, emitted due to impacting gas molecules. 

The two contrasting interpretations of the LP data have the commonality that they both identify Cassini as having charged to positive floating potentials. \citet{Zhang21} examined the role of charged dust in charging Cassini and showed that this could charge Cassini to positive potentials when the negatively charged ion/dust mass was significantly greater than that of the positive ions/dust and when the electrons constitute less than 10 \% of the total electron density. In this study we evaluate the hypothesis that secondary electron emissions in Saturn's ionosphere might have induced a similar effect. Given the differing interpretation of the LP data, we focus on evaluating the effect on the spacecraft potential to provide complementary understanding of the underlying system state. The dynamics of the spacecraft floating potential in these conditions are also relevant to low energy plasma measurements obtained during any high-velocity flyby missions of planetary and cometary environments. 

To study the effect of secondary electron and ion emission (SEE and SIE) for the Cassini spacecraft during the Grand Finale, we utilise three-dimensional particle-in-cell (PIC) simulations, as follows: Section 2 introduces the methods of the simulation and describes the input parameters. Section 3 analyses and discusses the results of the simulations as well as the varying of key parameters. Section 4 then concludes by summarising the results and discusses the implications to our understanding of the low energy plasma measurements of the composition of Saturn's ionosphere.

\section{Method}

In this study we utilise the three-dimensional Particle-In-Cell simulation code for ElectroMagnetic Spacecraft Environment Simulation (EMSES) developed for a self-consistent analysis of spacecraft-plasma interactions at electron scales \citep{Miyake09}. We embed a toy model of Cassini to scale within a predefined simulation domain. 
The three-dimensional simulations are run in the spacecraft frame where the inflowing ionospheric plasma consists of drifting Maxwellian velocity distributions. Each species has mass and charge normalized to the proton scale with a real ion-to-electron mass ratio. 
The spacecraft is treated as a perfect conductor, and a detailed description of conductors and the numeric can be found in the previous work \citep{Zhang21}.

We model Cassini at the representative altitude of 2500 km during Rev 292 as in the previous study \citep{Zhang21} but instead of including ``dust'' particles, we evaluate the hypothesis that SEE and SIE currents present a viable alternative to charged dust currents. 
To thus compare the effect of dust and secondary electron emissions accurately and independently, we use similar environment parameters as our previous study where the dust is investigated, but we replace the dust populations with secondary ion and electron emissions. 
We then scale our simulation parameters across multiple orders of magnitude to represent a larger range of Saturn's ionosphere, as explained below.

To test the hypothesis, we balance the electron populations to match the positive ion densities derived from the Langmuir Probe \citep{Morooka19} and introduce secondary electron and ion particles emitted from the spacecraft due to neutral-spacecraft collisions. The bulk current, I$_{total}$, onto Cassini can therefore be broken down into the electron current, I$_{electron}$, the ion current, I$_{ion}$, and the secondary currents I$_{SEE}$ and I$_{SIE}$:
\begin{equation}
      I_{total} = I_{electron} + I_{ion^+}  + I_{SEE}^{e} - I_{SEE}^{r} + I_{SIE}^{e} - I_{SIE}^{r}
      \label{totalI}
\end{equation}
where, importantly, I$_{SEE}^{e}$ and I$_{SIE}^{e}$ are the emitted electron and ion currents, respectively, and I$_{SEE}^{r}$ and I$_{SIE}^{r}$ represents those returning to impinge upon Cassini.
The same density of electrons and positive ions with mass 1.35 u are introduced in the simulations, as inferred from Langmuir Probe observations of the effective positive charge carrier at this altitude \citep{Morooka19}. Increasing the positive ion mass was also found to have only a small impact on the potential \citep[][Fig 5a therein]{Zhang21}. We also consider a cooler ionosphere of $370~K$. This temperature change is motivated by ionospheric models \citep{Moore2008,Moore18,Wodarg19,Yelle18} indicating ionospheric temperatures lower than the electron temperature inferred from Cassini's Langmuir Probe \citep{Morooka19} which is suggested to have been affected by secondaries \citep{Johansson22}.
As shown in the subsequent results, the differing temperature choices of \citet{Zhang21} and this study do not affect the trends reported. This choice also results in a smaller electron Debye length than previously considered. This requires a smaller grid width of $5~cm$ with a total grid of $256^3$, across a total simulation box size of $12.8~m^3$. 

The emitted SEE and SIE current densities are a function of the atmospheric neutral number density, $n$, elementary charge, $e$, the spacecraft velocity, v$_{sc}$,  the yield defined as the number of electrons ejected per incident neutral, $\gamma$, and the spacecraft's geometric cross-section, $A$, as:
\begin{equation}
    I_{SEE/SIE}^{e}  = \sum_{\alpha} \hspace{0.5mm} n_{\alpha}\hspace{0.5mm} e \hspace{0.5mm} v_{sc}  \hspace{0.5mm} \gamma_\alpha \hspace{0.5mm} A,
    \label{sumI}
\end{equation}
where $\alpha$ represents the neutral species of interest.  Due to the lack of laboratory experiments of quantum yields from Cassini's surface materials, the inclusion of emission requires careful consideration and as a result we utilise and vary yields associated with water ions. 

\citet{Schmidt85} determined yields experimentally from water molecules incident on three materials relevant for Giotto's { 70 km/s} flyby velocity of Comet 1P/Halley. Here we took the measured value of $\gamma$~=~0.15 as our base value of SEE impact yield in our study. At 2500~km altitude, these values correspond to $\approx 10~\mu A / m^2$. However, it is necessary to also vary this emission rate across a large range of values when applying this in the actual model, for two reasons. Firstly, there are significant uncertainties in adopting this rate for Cassini's Kapton blankets \citep{Lin95} and its lower {35} km/s flyby velocities compared to the Giotto's { 70 km/s} velocity that \citet{Schmidt85} did their experiment on. Secondly, the neutral density increases exponentially with decreasing altitude in Saturn's ionosphere \citep{Yelle18} and, as the secondary emission is directly proportional to the neutral density, varying the emission yield therefore captures this natural variation in Cassini's interaction with Saturn's ionosphere \citep{Moore18,Wodarg19}. Therefore, by varying the emission yield, $\gamma$,  across multiple orders of magnitude, we qualitatively recover a feature of the interaction between the Cassini spacecraft and Saturn's ionosphere that varies with altitude, from little to no neutral content at the top-side ionosphere down to the densest part of Saturn's ionosphere sampled. The variation therefore also qualitatively captures a variation of altitudes.
In this regard, the sensitivity of Cassini spacecraft potential to the emission density at different regimes, as we will see below, may also be of use for future high-velocity flyby missions of ionospheric environments.

The secondary ion emission is anticipated to be between 5--40 times lower than electron yields and we therefore implement this to be 10~\% of the electron yields with an emitted ion temperature of 10~eV \citep{Schmidt85}. 
In our simulations the electron and ion secondaries are emitted as a Maxwellian distribution. 
 We adopt yields due to neutral water ion density of $n$ = 1.5 $\times$ 10$^4$ cm$^{-3}$ from the model of \citet{Moore18} due to the significant effect of these species on the Giotto and Vega spacecraft but, as discussed previously, the variation in $\gamma$ can be viewed as interchangeable with variations in $n$ and therefore also altitude. Emission from further species such as CH$_4$ and CO$_2$ might also contribute given that kinetic electron emission processes \citep{Sternglass57} will dominate over potential emission ones \citep{Kishinevsky73} in this regime, but are not included at this stage.

\begin{table}[ht]
\caption{Environmental and System Simulation Parameters}
\centering
\begin{tabular}{|
>{}l 
>{}c |}
\hline
\multicolumn{2}{|c|}{\textbf{Environmental Parameters}} \\                   
Plasma ion density, $n_0$                            & 505  cm$^{-3}$ \\
Ion mass, $m_i$                                  & 1.35 amu                     \\
Electron temperature, $T_e$                      & 0.0318 eV (370 K)                      \\
Ion temperature, $T_i$                           & 0.0318 eV (370 K)                     \\
Magnetic field, $\vec{B}$                            & [1.48$\hat{x}$, --14.8$\hat{y}$, 1.24$\hat{z}$] $\mu$T                     \\
Flow velocity, $\vec{v}_{flow}$                             &  [--0.25$\hat{x}$, --32.4$\hat{y}$, --10.7$\hat{z}$] km s$^{-1}$                   \\
Ion acoustic speed, $v_{S}$             & 2.47 km s$^{-1}$                    \\
Debye length, $\lambda_D$                         & 5.90 cm                    \\
Electron gyroperiod, $\tau_{ge}$                   & 4.76 $\mu$s                     \\
Electron plasma period, $\tau_{pe}$                  & 4.98 $\mu$s                    \\
Ion gyroperiod, $\tau_{gi}$                        & 5.94 ms                     \\
Ion plasma period, $\tau_{pi}$                       & 0.247 ms                    \\
Electron emission density, $J_{SEE}$                    & 0.005 - 500 $~\mu A / m^2$ \\

Electron emission temperature, $T_{SEE}$                & 2~$eV$\\
Ion emission density, $J_{SIE}$                    & 0.0005 - 50 $~\mu A / m^2$ \\

Ion emission temperature, $T_{SIE}$                & 10~$eV$\\
\multicolumn{2}{|c|}{\textbf{System Parameters}}\\
Grid width, $\Delta$r                           & 5 cm                       \\
Time step, $\Delta$t                            & 0.033 $\mu$s                   \\
Simulation time, $t$                            & 0.67 ms                    \\
Particles per cell                        & 25

\\\hline
\end{tabular}
\label{table}
\end{table}

\section{Results \& Analysis}

\subsection{Plasma Interaction}

 \begin{figure}[ht]
           \hspace{-5mm} 
     \includegraphics[width=1.1\textwidth]{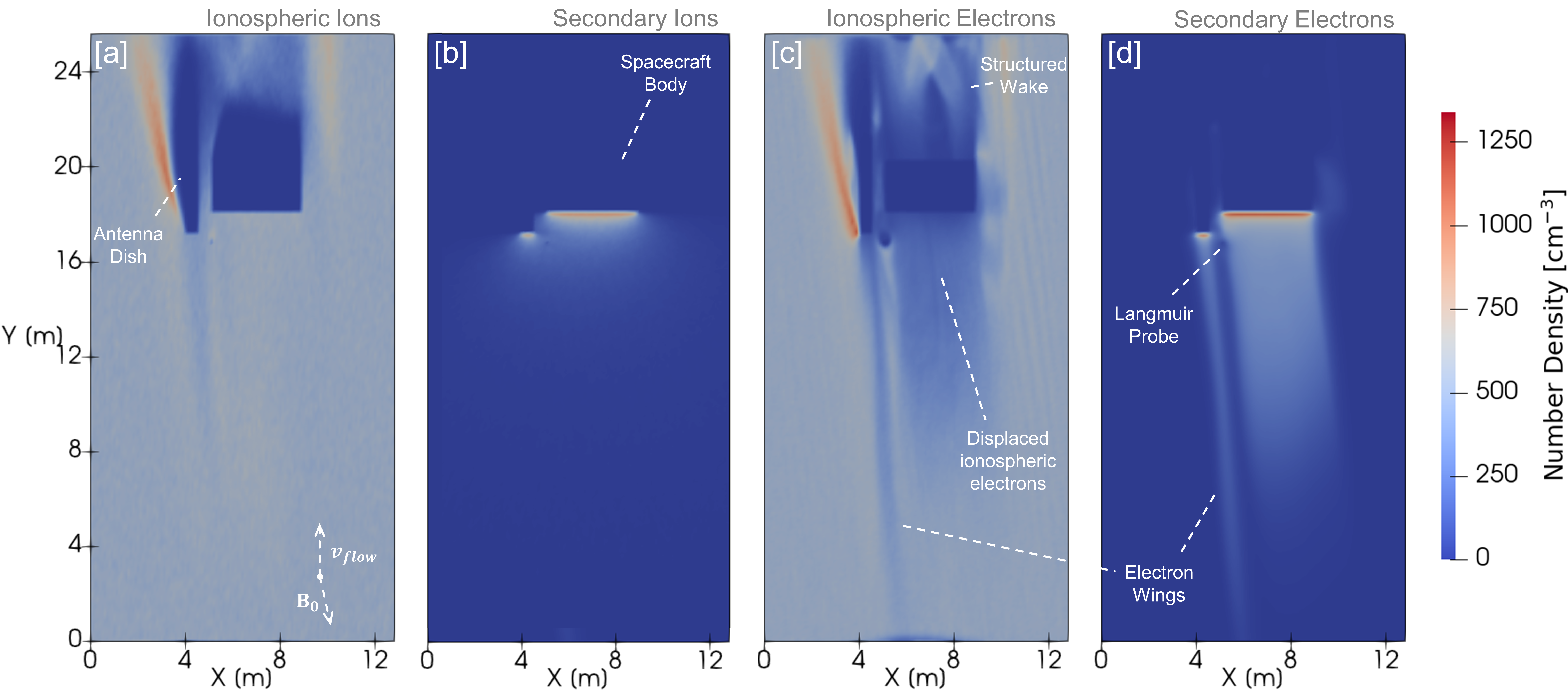}
     \caption{Simulated plasma densities around the Cassini spacecraft. [a] shows the ionospheric ions, [b] shows the secondary emitted ions, [c] shows the inospheric electrons, and [d] shows the secondary emitted electrons. The plasma velocity, v$_{flow}$, is predominantly along the Y-axis and the magnetic field, B$_0$ is approximately anti-parallel to this. Specific input parameters can be found in Table 1. The ionospheric electrons are {electrostatically} displaced upstream resulting in a combination of ionospheric and secondary electrons surrounding the spacecraft. Electrons wings caused by propagating Langmuir waves further modify the plasma ahead of the spacecraft and the negatively biased Langmuir Probe is visible as a region devoid of plasma. A schematic of the simulated spacecraft geometry can be found in \citet[][Figure 1 therein]{Zhang21}}
      \label{paraview}
 \end{figure}
 
 Figure \ref{paraview} shows the global plasma interaction between the Cassini spacecraft and Saturn's ionosphere at 2500 km altitude for the conditions outlined in Table \ref{table}. The colour-bar depicts the ion and electron densities (primary and secondary) and the plasma is moving along the positive y axis, and the magnetic field is approximately parallel to the -y axis. The spacecraft charges here to a positive potential.
 
 A plasma wake can be clearly seen trailing behind the spacecraft at regions where the density is depleted. Due to the high speed of the plasma flows and non-zero potential of the spacecraft, the incoming ions and electrons are deflected around the sides of the spacecraft forming enhanced densities adjacent to the wake. The probe swept from --4 to +4 V in Saturn's ionosphere and is negatively biased in this simulation, thus appearing as a region void of electrons. This biased potential also affects the surrounding plasma that subsequently impinges upon Cassini.

 Secondary electrons and ions are generated at the surfaces of impact simulating the incoming neutral impacts and result in electron and ion concentrations more than double the ambient ionospheric densities. These are generated in the spacecraft frame and are therefore able to diffuse away from the surfaces and upstream. This notably results in a decrease in the incoming ionospheric electron density ahead of the spacecraft. A unique aspect of this plasma regime is that the electron gyro-radii is smaller than the spacecraft while the ion gyro-radii is significantly larger. The emitted ions there appear to diffuse uniformly out in space upstream whereas the electrons are tied to the field lines. This notably presents a prediction for when they might be detected, i.e. for a geometry where the Langmuir Probe is magnetically connected to Cassini's main body or antenna dish.

 In front of the spacecraft, ``electron wings'' are present formed by Langmuir waves propagating along the background magnetic field \citep{Miyake20} which is oriented predominantly anti-parallel to the plasma flow. This appears to be notably enhanced compared to situations without SEE \citep{Zhang21} due to the enhanced electron densities resulting from SEE. Due to this effect striking the in-flowing boundary condition, the simulation box was expanded upstream as shown in Figure \ref{paraview} up to the point where the wing structures no longer intersected the upstream plasma. This verified that this effect produced negligible ($<$1 \%) differences in the simulation results.

 \begin{figure}[ht]
     \begin{center}
             \includegraphics[width=0.8\textwidth]{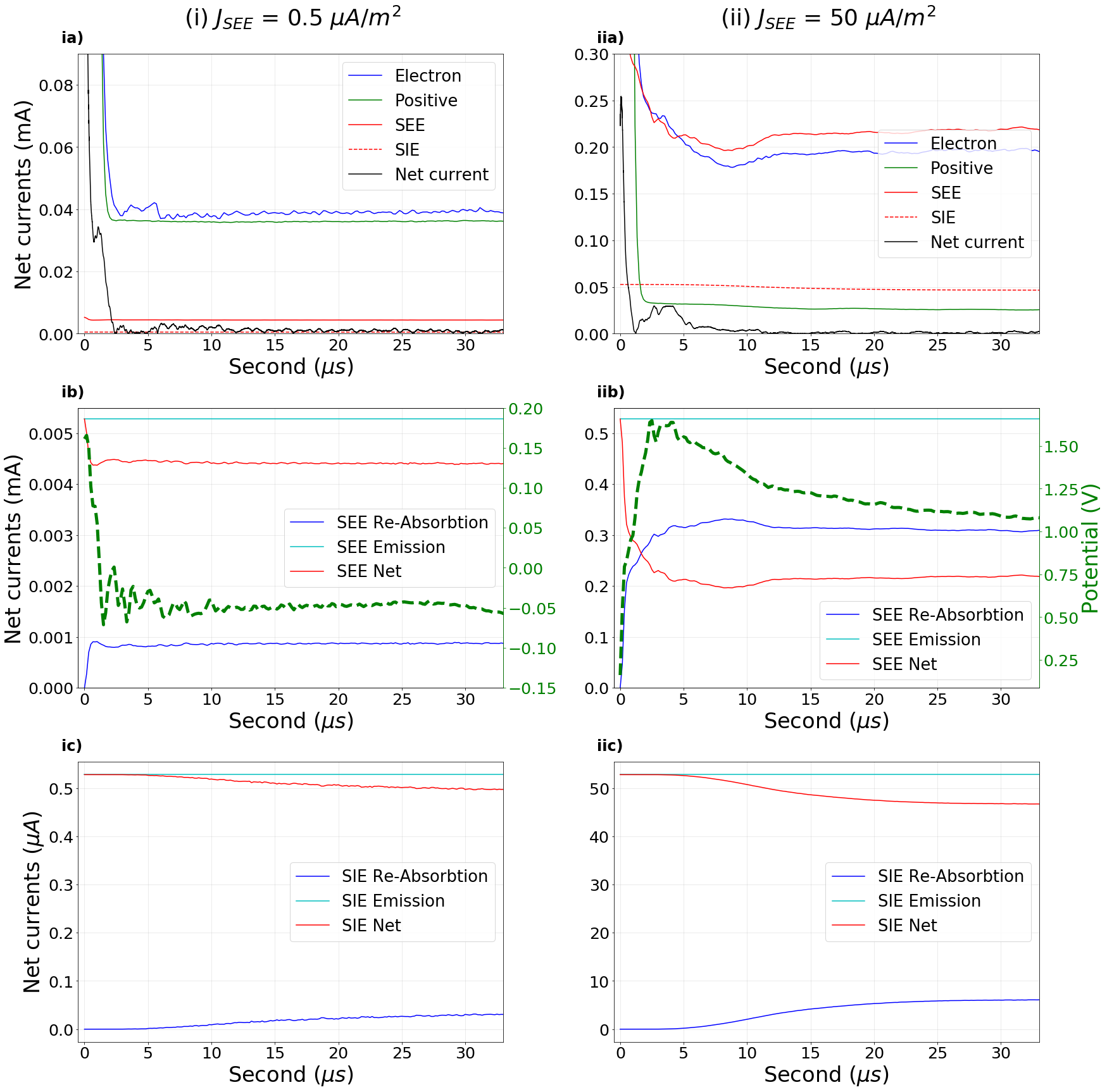}
     \end{center}
     \caption{Currents onto the Cassini spacecraft's for two distinct regions:  (ia--c) show the case of a negative floating potential induced when I$_{SEE}$= 0.5 $\mu$A/m$^2$ and I$_{SIE}$= 0.05 $\mu$A/m$^2$, and (iia--c) show the case of a positive floating potential induced when I$_{SEE}$= 50 $\mu$A/m$^2$ and I$_{SIE}$= 0.5 $\mu$A/m$^2$. The upper panels (ia) and (iia) show all the positive and negative currents inclusive, panels (ib) and (iib) show the currents associated with the emitted electrons along with the spacecraft potential and the lower panels (ic) and (iic) shows the currents associated with the emitted ions.}
      \label{currents}
 \end{figure}
 
 \subsection{Secondary Emitted Currents}
 
 Figure \ref{currents} shows the current decomposition comparison for spacecraft at relatively low secondary emission density (ia--c) and at high secondary emission density (iia--c). {These two scenarios represent two distinct regimes of Cassini charging to negative and positive potentials, as these simulations indicate occurs as Cassini descends into Saturn's ionosphere.} The ultimate values reached in Figure \ref{currents} are therefore of relevant to Cassini at Saturn while the time-history reveal the time-dependent interactions between the currents as the simulations reach steady state. 
 The most interesting result is that there is significant re-absorption of the emitted electrons back onto the spacecraft. In Figure \ref{currents}(iib), where the spacecraft is charging to a significant positive potential, over 50\% of the emitted secondary electrons are re-absorbed back onto the spacecraft later, resulting in the net yield of the SEE emission being less than 40\% of the neutral yield one would expect. Even in the case where there is little emission and the potential is negative, there is still significant re-absorption of the SEE electrons due to the space-charge-limited effect, i.e. the Child-Langmuir law. This makes the net effect of the SEE current environment dependent, as well as a diminishing return of SEE density when the spacecraft becomes significantly positive, as can be seen in the later Figures. The ions are however absorbed onto the spacecraft in much lower amounts, due to their larger emitted energies and greater momenta.

 For the high emission case, even though a majority of the SEE electrons are being re-absorbed, due to its high density it still dominates the positive currents in the system as its ``net'' current is still much larger than the positive ion currents in the system. As a result, the spacecraft's current balance and hence its potential is controlled largely by the properties of the SEE currents.

 In contrast, for the low emission case, as shown in Figure 2(ib), since the density is much smaller, the ``net'' SEE current is now much smaller and the positive ion current becomes the significant current in the system, hence in this case the spacecraft would not be sensitive to the properties of the SEE currents.

\subsection{Varying the Secondary Emission}

\begin{figure}[ht]
     \begin{center}
         \includegraphics[width=0.5\textwidth]{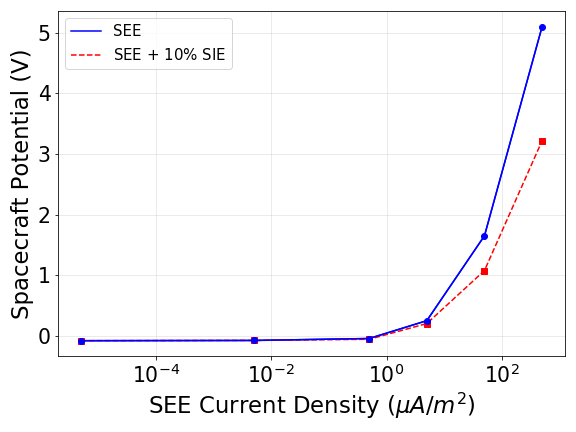}
     \end{center}
     \caption{Spacecraft potential dependence upon the secondary electron emission current density (yield) with zero ion emission (blue) and where the ion emission is 10 \% of the electron emission (red).} 
      \label{gamma}
 \end{figure}

Figure \ref{gamma} shows  the overall potential changes when one varies the secondary electron and ion emission currents in Equation \ref{sumI}. When the emission density is low ($<0.1~\mu A /m^2$) as expected at the top of the ionosphere, the potential of the spacecraft is virtually unchanged compared to when the secondary electrons and ions are not emitted. 
On the other hand, when the emission is high as expected for higher neutral densities lower in the ionosphere, the spacecraft potential becomes very sensitive to the emitted currents and, not only do they successfully bring the spacecraft potential to positive values, they are able to raise its potential to up to $>3~V$ at $500 \mu A /m^2$ SEE current density. This shows secondary currents, with SEE in excess of SIE, are indeed able to raise the spacecraft potential significantly, thus achieving some of the same global effect on the spacecraft as dust currents. 

We now compare our spacecraft potential results to Cassini measurements.
In the absence of SEE and SIE,  the simulated spacecraft potential is close to zero ($-0.08~V$) for the environmental conditions considered, as indeed is anticipated for an object moving through a cool, dense ionosphere. This baseline potential is dependent upon the electron temperature, if a temperature of 3 times higher is used, a starting negative potential of $-0.46~V$ will be obtained. This value is quite similar to the potential obtained in \citet{Zhang21}, where the same temperature but with dust included obtained a potential of $-0.42~V$. Cassini at this altitude measured at $-0.12~V$, close to the simulated environment with a cold plasma.

The variation in secondary emission density, as shown in Figure 3, therefore represents a clear departure to the potentials with clear dependence between the spacecraft potential and the neutral density, albeit mediated by the unknowns in the quantum yields. Using estimated yields and densities outlined Section 2, Cassini is estimated to experience $\approx 5\mu A /m^2 $ of SEE and SIE at higher altitude (240 0km) and $50\mu A /m^2 $ around the lowest altitude it experienced (1700 km). This corresponds to the range where SEE and SIE begin to make significant impact towards the potential of Cassini, as shown in the Fig 3. Although there is much uncertainties surrounding these estimations, this illustrated the possibility of SIE and SEE becoming a factor in the positive spacecraft potential during Cassini's flybys. The spacecraft potentials reported by the Langmuir Probe \citep{Morooka19} show variations from just below --1 V to +0.6 V using an estimate from the maximum  derivative of the current onto the probe. \citet{Johansson22}, however, suggest that the additional consideration of SEE changes the sweep interpretation and identifies higher potentials through determining the change between exponential and linear regions of the electron current in the current-voltage sweeps. The simulations presented herein therefore present constraints on the underlying system state, which can inform the various methods of inferring the spacecraft potential. 

\subsection{Emitted Electron Temperature}

\begin{figure}[ht]
     \begin{center}
             \includegraphics[width=0.5\textwidth]{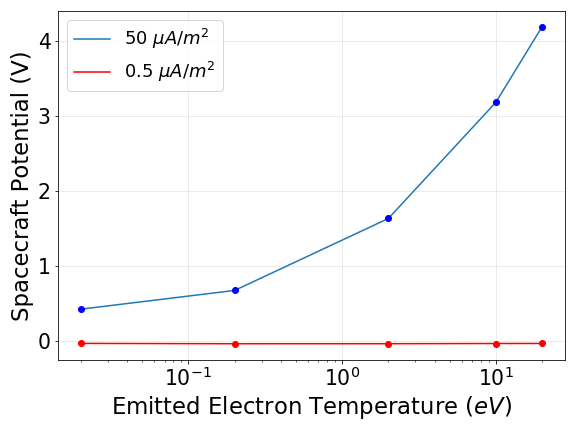}
     \end{center}
     \caption{Spacecraft potential dependence upon temperature of the emitted electrons.}
      \label{temperature}
 \end{figure}

As electron emission is anticipated to dominate over ion emission, further attention is given to the properties of the emitted electrons.
The temperature of the emitted electrons was implemented at $2~eV$, as anticipated by \citet{Schmidt85}, but this might well be be different for Cassini's surface's interaction with Saturn's ionosphere and \citet{Johansson22} indeed indicate a lower temperature of 0.5~eV. Figure \ref{temperature} therefore shows the sensitivity of the SEE current simulated under high and low emission current densities by varying the secondary electron's temperature. When the SEE current's magnitude is low, varying the temperature of the SEE species has almost no impact on the potential value of the spacecraft, and the spacecraft potential does not become positive. This is an expected result as, when the floating potential is negative, the electrons are strongly repelled and the emitted current density remains constant. However, for the  much higher $50~\mu A /m^2$ current density case, by raising the temperature of the electrons by a factor of $10$, the potential of the spacecraft raised from $1.6~V$ to almost $2.5~V$. 
This trend supports the analysis of Figure \ref{currents} by showing the higher the current density of the SEE electrons, the more sensitive the spacecraft's potential is to the emitted electron temperature. This therefore shows that when the spacecraft potential becomes positive, the characteristics of SEE might better provide an indicator of the characteristics of the neutrals striking the spacecraft.

The secondary electron emission temperature inferred by \citet{Johansson22} is $0.5~eV$, and notably lower than the laboratory derived rates of \citet{Schmidt85}. 
The variation of the SEE temperature in Figure \ref{temperature} covers the temperature inferred by \citet{Johansson22} under its varying range. The resultant trend showed that the spacecraft potential varies smoothly with emitted electron temperature when positively charged, and the spacecraft stays positively charged for large emission rate even at very low temperatures of $0.01~eV$. Therefore, a sufficient emitted electron current would theoretically drive the Cassini spacecraft to positive potentials in Saturn's ionosphere, as ~\citet{Johansson22} infers. 
This variation with temperature also allows these results to be applicable to future missions where the environment and emitted electron characteristics could be different.

\section{Discussion \& Conclusions}
In conclusion, we use three-dimensional PIC simulations to demonstrate that SEE theoretically represents a phenomenon for producing positive spacecraft potentials in Saturn's ionosphere. Specifically, when the amount of SEE is large ($>$1 $\mu A/m^2$), spacecraft potentials were very sensitive to the SEE yield and hence the simulations could produce a smooth transition from negative to positive values as observed during the Grand Finale flybys \citep{Johansson22,Morooka19}. { For small SEE yields, however, SEE induced a negligible effect on the simulated Cassini spacecraft's plasma interaction.}

The simulations show the emitted electrons propagate upstream of the spacecraft along the magnetic field and can then be re-absorbed, which means they might also be detected by the Langmuir Probe and other plasma instruments for specific spacecraft-magnetic field orientations. Figure \ref{paraview}  highlights how measurements of the ionospheric electrons might also be affected by the production of secondary electron populations. Identifying these re-absorbed electrons could also be useful for identifying SEE populations by other instruments on-board Cassini, as well as helping to calibrate for Langmuir Probe analysis of the ionospheric content. 

The inference of charged dust populations in Saturn's equatorial ionosphere \citep{Morooka19,Wahlund18} and the charge depletion of electrons of over 90\% is consistent with Langmuir Probe observation at Enceladus \citep{Morooka11,Wahlund09} and Titan \citep{Agren09,Shebanits16}, where large negatively charged ions and dust had been detected using Cassini's Plasma Spectrometer \citep{Coates10,Coates07,Desai17a,Wellbrock19,Mihailescu20}. At Saturn, the presence of negatively charged ions and dust is explained through the accumulation of in-falling ring particles \citep{Hsu18,Mitchell18} which undergo electron impact ionisation processes.
In a preceding study, \citet{Zhang21} thus showed that charged dust can also produce a positive spacecraft potential when the positive species are overall more mobile than the negative species,with electron depletions of over 90 \%. This study, however, shows this is potentially explained by the phenomenon of neutral-induced electron and ion emission with electron emission rates dominating over the ion emission rates.

The amount of dust outside of the ionosphere is constrained by both CDA \citep{Hsu18} INCA/CHEMS \citep{Mitchell18}, and RPWS \citep{Wahlund18} and the amount of dust that falls into the equatorial region ionosphere from above D-ring is estimated at around 10--100 cm$^{-3}$ at 1500 km as projected by models to lower altitudes. A major outstanding question therefore remains as to the fate of these inflowing dust populations. When considering only the spacecraft potential, the two effects of SEE and charged dust cannot be distinguished from one another, and it is possible that both contributed to the positive potential observed at Saturn.

A definitive question within the SEE debate is what emission yields to use for which neutral species incident upon Cassini which highlights the urgent need for further laboratory studies therefore. Here we used emission rates typical for metals and  and those closest to the conditions at hand \citep{Schmidt85} but these are still not directly representative of Saturn's atmospheric neutrals impacting Cassini as they were designed for the significantly higher, {70 km/s}, velocity flybys of the Giotto and Vega missions \citep{Grard89} {compared with Cassini's 35 km/s velocity}. In this study we also only considered water molecule densities derived from the ionospheric model of \citet[][Figure 2 therein]{Moore18}. Observations from the Grand Finale revealed significant populations of  methane, ammonia, and organics in addition to the anticipated molecular hydrogen, helium, and water \citep{Hsu18,Mitchell18,Waite18} and NH$_3$, CO$_2$, CH$_4$ and the ambiguous mass 28 detections, all having densities the same or higher as H$_2$O. These should all have energies sufficient to trigger electron emission from Cassini as their energies in the spacecraft frame are expected to exceed the work function of the target surfaces. If these species have similar yields, the SEE current densities should be several factors, if not an order of magnitude higher which seems unphysically large. Such elevated SEE currents would drive Cassini to even higher potentials but as the potential exceeds the peak of the Maxwellian of the emitted electron energies, fewer and fewer would be able to escape the potential well surrounding the spacecraft \citep{Marchland14}. The emitted ions would, however, easily escape due to electrostatic repulsion. In this scenario the potential would apparently be mediated by the SIE currents which act to prevent the potential diverging to extreme positive potentials.

A more accurate Cassini spacecraft model could also be used within further studies. For example, while the spacecraft is generally designed to be conducting, the high-gain-antenna is coated in a resistive paint, the properties of which are not included herein. The most important factor for the plasma interaction is however the ram-pointing side of Cassini and so the first Grand Finale flyby, where Cassini flew with the HGA in ram, this effect would be most important.
Given the uncertainties in the measurements of Saturn's ionospheric plasmas and the multitude of parameters that might therefore be varied, 
we have therefore opted to sweep through the most important parameters of interest. We directly compared our potentials, Figure 3 and Figure 4, and found potentials similar to those inferred for Cassini \citep{Morooka19,Johansson22} and the simulations results and trends discovered are therefore applicable to studies of spacecraft charging in Saturn's ionosphere and in similar environments.

It is also worth noting that if accurate quantum yields were determined from neutral molecules onto Cassini thermal Kapton blankets, the spacecraft potentials might also yield further information on the neutral composition of the giant planet's atmospheric densities as the remaining unknown in Equation \ref{sumI}.
In future missions where dust effects are small, accurate quantum yields
might therefore be used to infer information on neutral populations from the spacecraft potential and measured incident currents. 
\section*{Acknowledgements}

ZZ acknowledges funding from the Royal Astronomical Society. RTD acknowledges STFC Ernest Rutherford Fellowship ST/W004801/1 and
NERC grants NE/P017347/1 and NE/V003062/1. YM and HU acknowledge
grant no. 20K04041 from the Japan Society for the Promotion of
Science: JSPS, and support from the innovative High-Performance-Computing Infrastructure (HPCI: hp210159) in Japan. OS acknowledges SNSA grant no. Dnr:195/20. FLJ acknowledges a grant from Lennanders stifelse.
This work used
the Imperial College High Performance Computing Service (doi:
10.14469/hpc/2232).

\bibliographystyle{aasjournal}
\bibliography{main}

\end{document}